\documentclass[english]{article}
\usepackage[T1]{fontenc}
\usepackage[latin9]{inputenc}
\usepackage{geometry}
\usepackage{color}
\usepackage{array}
\usepackage{textcomp}
\usepackage{amstext}
\usepackage{amssymb}
\usepackage{graphicx}

\makeatletter

\newcommand{\lyxmathsym}[1]{\ifmmode\begingroup\def\b@ld{bold}
  \text{\ifx\math@version\b@ld\bfseries\fi#1}\endgroup\else#1\fi}

\providecommand{\tabularnewline}{\\}

\makeatother

\usepackage{babel}
\begin{document}
\title{Implementation of coherent one way protocol for quantum key distribution
up to an effective distance of $145$ km}
\author{Priya Malpani$^{1}$, Satish Kumar$^{2}$, Anirban Pathak$^{3}$ \\
 {\small{}Jaypee Institute of Information Technology, A-10, Sector-62,
Noida, UP-201309, India}\\
 $^{1}$priya.ims07@gmail.com, $^{2}$mr.satishseth@gamil.com, $^{3}$anirban.pathak@gmail.com,}
\maketitle
\begin{abstract}
It's well known that many of the traditional cryptographic schemes
(like RSA and DH) will be vulnerable when scalable quantum computers
will be available. However, quantum cryptographic protocols can provide
unconditional security. One such protocol for unconditionally secure
quantum key distribution (QKD) is coherent one way (COW) protocol.
In the present work, we report experimental realization of an optical
fiber based COW protocol for QKD in the telecom wavelength ($1550$
nm) where the attenuation in the optical fiber is minimum. A laser
of $1550$ nm wavelength, attenuator and intensity modulator is used
for the generation of pulses having average photon number $0.5$ and
repetition rate of $500$ MHz. The experiment is performed over $40$
km, $80$ km and $120$ km of optical fiber and several experimental
parameters like disclose rate, compression ratio, dead time and excess
bias voltage of the detector are varied for all the cases (i.e., for
$40$ km, $80$ km and $120$ km distances) to observe their impact
on the final key rate. Specifically, It is observed that there is
a linear increase in the key rate as we decrease compression ratio
or disclose rate. The key rate obtains its maximum value for least
permitted values of disclose rate, compression ratio and dead time.
It seems to remain stable for various values of excess bias voltage.
While changing various parameters, we have maintained the quantum
bit error rate (QBER) below $6$\%. The key rate obtained is also
found to remain stable over time. Experimental results obtained here
are also compared with the earlier realizations of the COW QKD protocol.
Further, to emulate key rate at intermediate distances and at a distance
larger than $120$ km, an attenuator of $5$ dB loss is used which
can be treated as equivalent to $\thicksim25$ km of the optical fiber
used in the present implementation. This has made the present implementation
equivalent to the realization of COW QKD upto $\thicksim145$ km.
\end{abstract}

\section{Introduction}

Sharing confidential information has always been a very challenging
task. The art of sharing the messages from one person/device to another
person/device is referred to as communication. The process of sending
secret messages which can only be decoded by a legitimate party is
known as cryptography. Cryptography is almost as old as human civilization
as the demand for hiding information from illegitimate party was in
existence from the beginning of human civilization. Initially, simple
tricks of sending secret messages were used, but with time the schemes
of hiding information evolved and many complicated protocols of cryptography
appeared. In classical cryptography, all the modern protocols for
cryptography are based on computationally hard problems. Lucidly speaking
in the heart of any modern classical cryptographic scheme, there is
a computationally hard problem that an adversary is not expected to
solve within the time frame for which we wish to keep an information
secure. Well known examples of such protocols are DH and RSA schemes.
The security of such schemes is conditioned on the computational power
of the adversary and no classical scheme can provide unconditional
security, but in sharp contrast to that there are schemes for quantum
key distribution (QKD) which can provide unconditional security. Further,
there are quantum algorithms which can solve classically hard computational
problems in polynomial time and thus ensure that several schemes of
classical cryptography, including RSA and DH will be vulnerable when
scalable quantum computers will be built. Here it to be apt to note
that if a key is secure then the message encoded with the key is also
secure and thus for a secure communication it's enough to distribute
a key in a secure manner. Consequently, most of the efforts in quantum
cryptography are restricted to design, analysis and implementation
of the schemes for QKD. 

In the background of the above, several protocols for QKD have been
proposed. The first of those was proposed by Bennett and Brassard
in 1984 \cite{BB84} and it drew considerable attention of the community
since it was able to provide unconditional security (in the sense
that security of it (and also of other schemes for QKD) followed from
laws of nature and was not conditioned on the computational power
of the adversary) which is a desired feature, but not achievable in
the classical world. Since then, several protocols for QKD \cite{bennett1992quantum,bennett1992quantumm,bruss1998optimal,ekertt1992quantum,hwang2003quantum,scarani2004quantum,lo2012measurement}
and other cryptographic tasks \cite{zhu2017experimental,sisodia2021optical}
have been introduced. Many of the proposed protocols are theoretical
in nature and never been implemented. There are only a handful of
schemes for QKD which have usually been experimentally realized. For
example, BB84 \cite{BB84,bennett1992experimental}, Coherent one way
(COW) \cite{stucki2005fast,Gisin2009COW}, Differential phase shift
(DPS) \cite{inoue2002differential,takesue2007quantum} are some of
the schemes which have been realized by various groups because of
the ease of implementation with the presently available technology
and because of the possibility of commercial deployment. Here, we
aim to implement COW protocol in a lab environment and analyze the
role of various experimental parameters on the performance of the
implemented protocol as reflected through the key rate (KR).

\begin{figure}
\centering{}\includegraphics[scale=0.6]{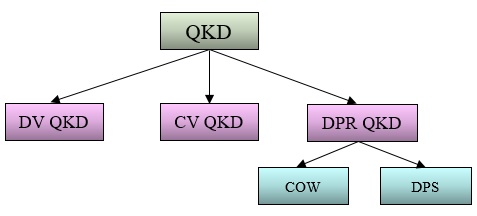}\caption{\label{fig:QKD_structure}(Color Online) Figure shows the classification
of QKD schemes. }
\end{figure}

Existing protocols for QKD differ from each other in many ways, and
they may be classified in various ways. Based on the nature of the
resources used for encoding operation they may be classified in 3
classes \cite{wang2019realistic} as (i) discrete variable QKD (DVQKD)
protocols \cite{BB84,bennett1992quantum,bruss1998optimal,hwang2003quantum},
(ii) continuous variable QKD (CVQKD) protocols \cite{ralph1999continuous},
and (iii) distributed phase reference QKD (DPR QKD) protocols \cite{inoue2002differential,stucki2005fast}
(see Fig. \ref{fig:QKD_structure}). DPR QKD protocols are unique
in the sense that at the measurement end, joint measurement is performed
on subsequent signals. Further, DPR QKD schemes can support secure
long distance QKD. In addition, DPR QKD implementation is robust against
polarization fluctuations as their coding of $1$ and $0$ is independent
of polarization degree of freedom \cite{Gisin2009COW}. Another advantage
of DPR protocols is that they are easy to implement experimentally
with the available technologies. Two schemes which belong to the class
of DPR QKD protocols and experimentally realized many times are COW
protocol \cite{stucki2005fast} and DPS protocol for QKD \cite{inoue2002differential}.
As mentioned above, here we aim to experimentally realize COW protocol
for QKD. Before, we describe the protocol and the setup built to realize
it, it will be apt to recall a few relevant exiting results related
to the COW QKD scheme. 

In a pioneering work reported in 2005, Stucki et al. introduced the
idea of COW protocol for QKD \cite{stucki2005fast}. In this protocol,
the secret key is shared between sender and receiver based on the
time of arrival of the information sent by the sender. The sender
and the receiver are connected by two channels, one of them is quantum
channel and the other one is a classical channel. Quantum channel
is used to share quantum signals and the second one is an authenticated
classical channel which is used for exchanging the time of arrival
information and for carrying out the information reconciliation to
achieve the final secret keys. This protocol has several advantages
like the ease of experimental implementation, its robustness against
photon number splitting (PNS) attack, it has tolerance against reduced
visibility, also it is polarization insensitive. Various research
groups have reported the security proofs for COW QKD against general
and coherent attacks \cite{gao2022simple,moroder2012security}. Being
free from PNS attack, this protocol is implementable using attenuated
light source which is an advantage as generation of single photon
on demand is an extremely challenging task with present experimental
resources. Further, upper bounds on the security of COW and DPS protocols
and two modified COW QKD protocols for distances $\geq50$ km have
been reported in \cite{branciard2008upper}. 

As noted above COW protocol has potential for commercial deployment
using presently available technology and is secure and the relevant
bounds are known. These facts, and the fact that it can be implemented
in lab level with very limited resource (like with only one single
photon detector is sufficient for its implementation in lab where
monitor line (to described later) can be ignored) have motivated us
to implement this protocol in our lab with an aim to see impacts of
various experimental parameters on the KR. Specifically, we have implemented
COW QKD over a single mode fiber starting with $40$ km distance and
extending up to $120$ km for various parameters like disclose rate
(DR), compression ratio (CR), excess bias voltage (BV) and detector
dead time (DT) maintaining QBER below $6\%$ during the whole experiment.
We have also emulated the effect of experimental parameters on the
KR at intermediate distances and at a distance larger than $120$
km by using an attenuator of $5$ dB loss which can be treated as
equivalent to $\thicksim25$ km of the optical fiber as the loss in
optical fiber used here is $0.2$ dB/km. It may be noted that the
experiment is realized in lab, KR obtained in lab is usually better
than that obtained during the field deployment. This happens because
of various reasons including low noise conditions of the lab. Certain
changes are usually made in the setup before field deployment. For
example, in \cite{stucki2009continuous}, InGaAs single photon detectors
(InGaAs SPDs) were used for lab level implementation and superconducting
nanowire single photon detectors (SNSPDs) which are costly but much
more efficient were used in field implementation. We will also show
that with the increase in distance, a bit of fluctuations on KR appears
for certain experimental parameters and consequently much care is
needed in selecting experimental parameters for the realization of
DPR QKD schemes for distance >$100$ km. 

The rest of the paper is structured as follows. In Section \ref{sec:COW-protocol-definition},
COW QKD protocol is briefly described and the experimental setup used
by us to implement it is explained. In Section \ref{sec:Steps-involved-in},
we have explained the steps involved in the calculation of secret
KR, the results are shown in detail in Section \ref{sec:Observation},
where we have reported stability of KR and its variations with respect
to various experimental parameters. In Section \ref{sec:Comparison},
we have critically compared the present implementation with the existing
realization of COW protocols for QKD. Finally, the paper is concluded
in Section \ref{sec:Conclusions}.

\section{COW protocol and the experimental setup used for its realization
\label{sec:COW-protocol-definition}}

\begin{figure}
\begin{centering}
\begin{tabular}{c}
\includegraphics[scale=0.5]{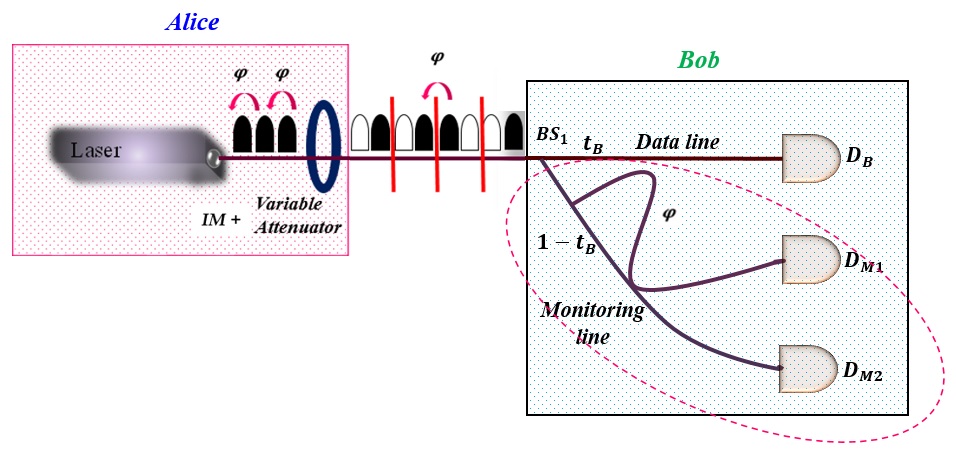}\tabularnewline
\end{tabular}
\par\end{centering}
\caption{\label{fig:COW}(Color Online) Illustration of COW protocol, in the
work reported here we have used only one detector (so monitoring line
is not implemented). However, to ensure that the presence of monitoring
line will not affect the KR obtained here, $BS_{1}$ shown above which
is a $90:10$ beam splitter is used in our setup. }
\end{figure}

\begin{figure}
\begin{centering}
\begin{tabular}{c}
\includegraphics[scale=0.5]{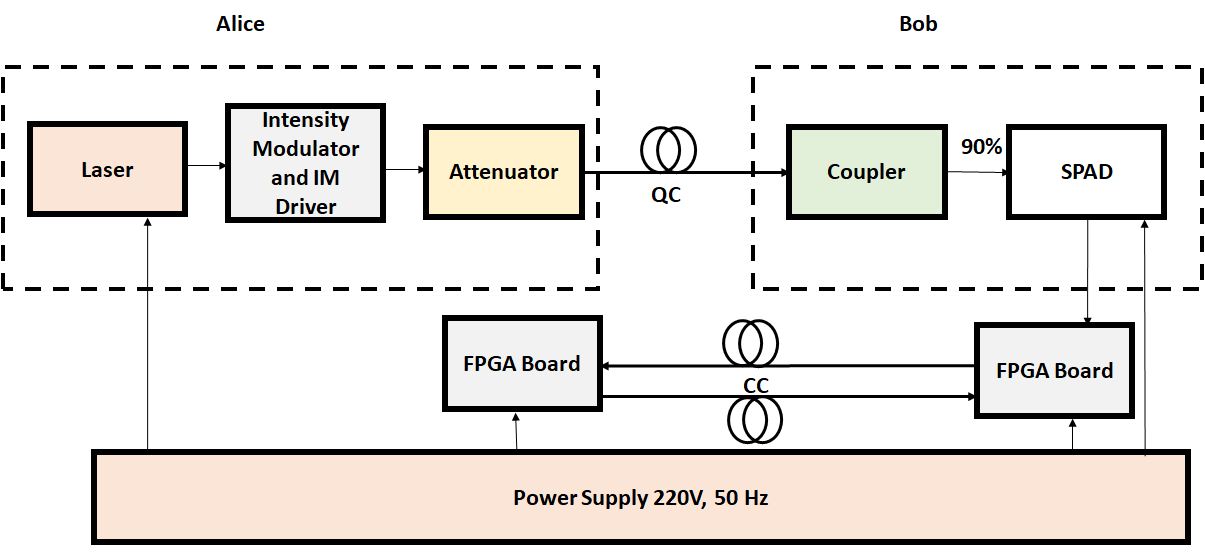}\tabularnewline
\end{tabular}
\par\end{centering}
\caption{\label{fig:COW-1}(Color Online) Fig shows the schematic of COW protocol,
QC is quantum channel and CC is classical channel.}
\end{figure}

\begin{figure}
\begin{centering}
\begin{tabular}{c}
\includegraphics[scale=0.5]{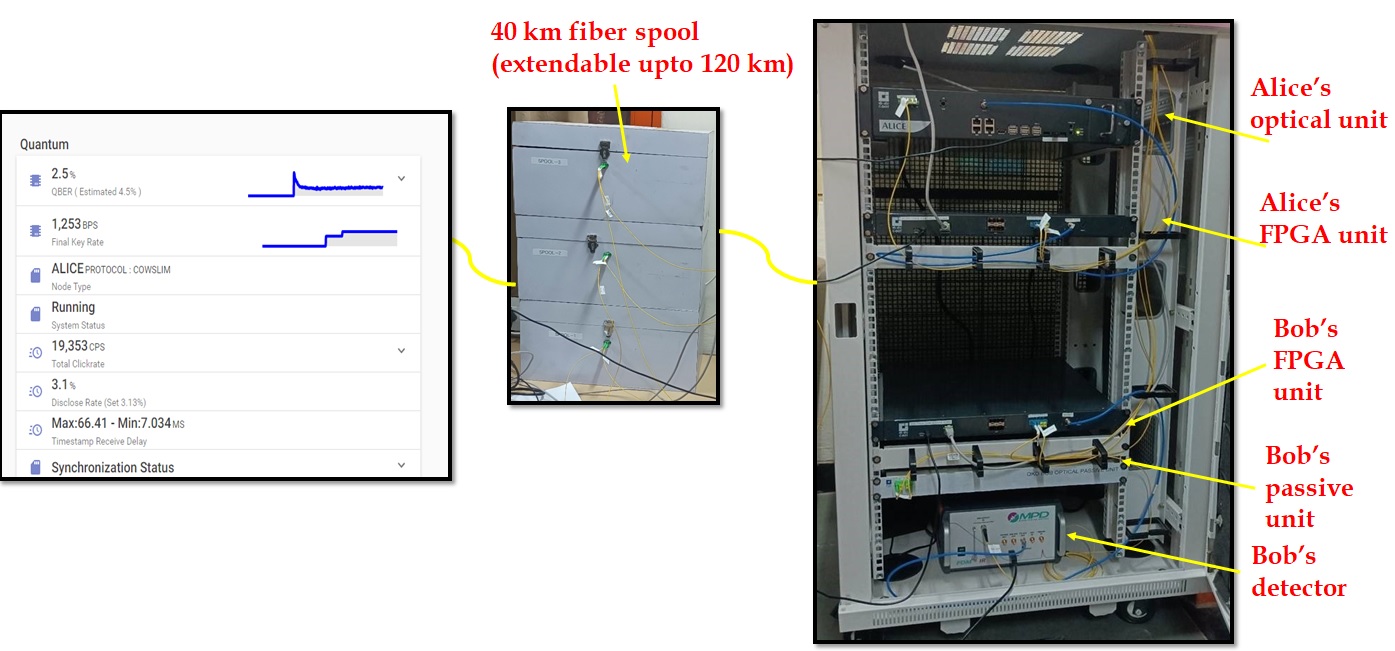}\tabularnewline
\end{tabular}
\par\end{centering}
\caption{\label{fig:COW-1-1} (Color Online) Lab setup for COW QKD protocol.
The right most unit is the actual setup of COW QKD protocol, middle
one is the collection of fiber spools and the left most is the GUI
for the same.}
\end{figure}

As mentioned above, COW protocol belongs to the class of QKD protocols
referred to as DPR QKD protocols. In these kind of protocols, the
security is derived from the coherence of sequential pulses. In COW
protocol, the encoding of a bit is done using a pair of empty and
non-empty pulse. The COW QKD protocol introduced by Stucki et al.
\cite{stucki2005fast} can be described in step-wise manner as follows:
\begin{enumerate}
\item In the first step, Alice prepares a sequence of pulses $\left|0\right\rangle \left|\alpha\right\rangle $(empty,
non-empty), $\left|\alpha\right\rangle \left|0\right\rangle $ (non-empty,
empty) and $\left|\alpha\right\rangle \left|\alpha\right\rangle $
(non-empty, non-empty) $\left(\left|\alpha\right|^{2}<1\right)$ corresponds
to logical bit $1$, $0$ and decoy respectively using attenuated
light source and intensity modulator with each logical bit having
probability $\frac{\left(1-f\right)}{2}$ and decoy with probability
$f$ and delivers the pulse sequence to Bob.
\item While receiving the pulse sequence, Bob measures the time of arrival
of $90\%$ photons on his detector $D_{{\rm B}}$ for the generation
of raw key and rest $10\%$ of photons on the monitoring line for
security purpose (refer to Fig. \ref{fig:COW}).\footnote{$90:10$ distribution of photons among data line and monitor line
mentioned here and implemented by us is standard. However, some groups
have implemented with $95:5$ distribution \cite{shaw2022optimal}.
Such an approach would lead to higher KR at the cost of security as
the number of pulses in monitor line will considerably reduces leading
to a possibility that an eavesdropping effect remains unnoticed.} 
\item Bob randomly checks the coherence between the non-empty pulses using
detector $D_{{\rm M1}}$ and $D_{{\rm M2}}$ and the set up of Mach
Zhender is in such a way that always $D_{{\rm M1}}$will click if
there is no disturbance by Eve (refer Fig. \ref{fig:COW}).
\item After that Alice and Bob perform sifting (removal of decoy sequences),
error correction, privacy amplification to obtain the secret key.
\end{enumerate}
For the experimental realization of the above described protocol,
a system has been developed in cooperation with CDOT \cite{Cdot}
in a rack mountable in four $19$ inches boxes (Shown in the extreme
right side of Fig. \ref{fig:COW-1-1}) which may be referred to as
CDOT system for convenience. The rack contains both Alice's station
and Bob's station. Apparently, they are in the same place in the lab,
but actually Alice are Bob are separated by long optical fiber. In
fact, two spools of fibers are used to link Alice and Bob (see middle
panel of Fig. \ref{fig:COW-1-1} where the spool of the optical fiber
of appropriate length is kept in the wooden boxes shown). One of the
fiber spools is used as the quantum channel between Alice and Bob
and the other one used as classical communication channel between
Alice and Bob. Each wooden box (equivalently spool of fiber) is being
connected to the ethernet for the classical communication and raw
key processing. Here, we are doing an experiment (schematic is shown
in Fig. \ref{fig:COW-1}) without considering Eve (i.e., without implementing
the monitoring line) as we are working in the lab environment with
a limited number of detectors. However, to ensure that the obtained
KR remains practical, $90:10$ optical coupler (equivalently shown
as beam splitter in Fig. \ref{fig:COW}) is used and only 90\% photons
are transmitted. In this situation, the probability of the random
sequence of logical bits $0$ and $1$ created by Alice would be half
($f=\frac{1}{2}$).  Here, we are using a classical true random number
generator (TRNG) for the generation of a sequence of pulses randomly
on Alice's side. The reason of using TRNG is that it is faster than
quantum random number generator (QRNG). However, in practice a hybrid
approach may be used where a QRNG will be used to provide seeds for
the pseudo random number generator (PRNG). Such an approach is already
realized in \cite{stucki2009continuous}, and technically it's a straightforward
modification. 

We have demonstrated COW protocol and key generation using optical
fibers. The experimental setup is shown in Fig. \ref{fig:COW-1-1}.
Alice uses a laser having telecom wavelength ($1550$ nm), attenuator
and intensity modulator for the generation of pulses having repetition
rate $500$ MHz. With the help of attenuator and modulator the pulses
are formed in such a way that $0.5$ photon per pulse is sent to Bob.
The experiment is performed over $40$ km, $80$ km and $120$ km
distance. Bob uses $90:10$ optical coupler to split the pulses. $90\%$
photons go to the detector along the data line for raw key generation
and the $10\%$ photons go to monitoring line for security check or
to check the presence of Eve. We aimed to implement it within lab
level with minimum resources and to do so, in this lab level experiment,
we have not implemented monitoring line as that allowed us to implement
COW protocol with one detector only instead of three required for
the full phased implementation or any implementation outside lab\footnote{An implementation of COW protocol with monitoring line and also an
implementation of DPS protocol in our lab will be reported separately. }. Here, in absence of Eve, $10\%$ pulses are considered as loss.
This 90:10 distribution of pulses is standard as mentioned above.
However, in \cite{shaw2022optimal} $95:5$ coupler is used means
$95\%$ photons are given to data line which contribute in key formation
and only $5\%$ photons will be used to monitor Eve's presence. This
approach will automatically increase KR at the cost of security. The
KR to be obtained here is practical. This is interesting because,
implementation of DPS would require at least 2 such detectors and
the detectors are costly. In a lab scenario where we are sure of Eve's
absence, it's a reasonable approach for analysis of KR and impact
of several experimental parameters on KR. The detector used at Bob's
side in our experiment is from MPD (PDM-IR). The DT for the detector
is varied from $50$ $\mu$s to $20$ $\mu$s. For classical communication
between Alice and Bob, FPGA boards are being used. In post processing,
the error correction technique used here is LDPC error correction
code and the privacy amplification is done using FFT based Toeplitz
hashing scheme, Table \ref{tab:System-component} shows the specification
of the components used in the experiment.

\begin{table}
\begin{centering}
\begin{tabular}{|c|c|}
\hline 
Component & Property/ Type\tabularnewline
\hline 
Laser & (Thorlab-SFL1550P) $1550$ nm\tabularnewline
\hline 
Fiber & SMF-$28$ (ITU-TG652D) (loss-$0.2$ dB/km)\tabularnewline
\hline 
Intensity modulator & Lithium Niobate based\tabularnewline
\hline 
Random number generator & TRNG\tabularnewline
\hline 
Operating Temperature & $10^{\lyxmathsym{\textdegree}}-3\ensuremath{0^{\lyxmathsym{\textdegree}}}$\tabularnewline
\hline 
Detector & SPAD (MPD-PDM IR)\tabularnewline
\hline 
Error correction technique & LDPC\tabularnewline
\hline 
Privacy amplification & FFT based Toeplitz hashing scheme\tabularnewline
\hline 
\end{tabular}
\par\end{centering}
\caption{\label{tab:System-component}Specifications of the components used
in our experiment.}
\end{table}

\section{Steps involved in estimating key rate\label{sec:Steps-involved-in}}

Distance and KR are two main parameters to evaluate an implementation
of QKD. Specifically, without compromising with security we wish to
achieve maximum possible distance and KR. Calculation of distance
is straightforward. In our case, it's the length of the optical fiber
used as a channel (say, the length of the optical fiber constituting
the quantum channel). Of course, we can add an attenuator causing
the fixed amount of loss and consider that as equivalent to a fixed
length of fiber assuming that unit length of optical fiber causes
a fixed amount of loss (in our case, $0.2$ dB/km). Calculation of
KR is not that straightforward and involves many steps, like the calculation
of counts, windowing operation, DR incorporation, factor of error
correction and privacy amplification. Total counts are counts of those
pulses which are received at Bob's side. Various factors related to
the quality of the channel, source and detector affect the total counts.
Specifically, to name a few important factors, we may mention the
loss in optical fiber, detector efficiency, pulse rate, the gap between
pulses and DT of the detector. For example, in the present implementation
the loss of fiber is $0.2$ dB/km. For $80$ km, the loss is $=0.2\times80=16$
dB \footnote{some more losses will happen because of macrobending as we are using
spool of optical fibers.}. Total received photons per pulse at Bob's detector side after passing
through fiber $=\frac{0.5}{10^{1.6}}=0.0126{\rm photon/pulse}.$ This
is so because $N_{out}=N_{in}\times10^{-\frac{{\rm loss}}{10}}$,
and in our case $N_{in}$ is $0.5$ photon/pulse. Since we are using
$90:10$ optical coupler therefore the data line receives $90\%$
of photons which gives $0.0126\times0.9=0.01134$ photons per pulse
in the data line. Now the detector used in our experiment has efficiency
of $10\%,$ so total number of photons/pulse received at the detector
of Bob would be $=0.01134{\rm \times0.1=0.001134{\rm p{\rm hoton/pulse.}}}$
Pulse rate of the laser is $500$ MHz so the total count $=500\times10^{6}\times0.001134=567000$.
The gap between the pulses is $\frac{1}{567000}=1.8{\rm \mu s}$ and
we may consider that the DT of detector in a specific set of experiment
is $50{\rm \mu s}$. Total time=Gap between pulses $+$ DT of detector
$=51.8{\rm \mu s}$. The total clicks are $=\frac{1}{51.8\times10^{-6}}=19305$.

Next, we need to estimate both QBER and secure key rate in the parameter
estimation step. In order to estimate QBER, a fraction of Alice's
and Bob's raw keys are required to be compared. Clearly, DR quantifies
the fraction of the raw keys which are compared to obtain QBER. After
computing, QBER on has to correct erroneous keys by applying suitable
error correction protocol. This process (error correction) is needed
to ensure that Alice and Bob share identical keys. Here, we have used
LDPC error correction protocol to correct erroneous keys. Once the
erroneous keys are corrected then the next step is privacy amplification\footnote{Privacy amplification plays a crucial role in any QKD scheme. The
role of privacy amplification is to provide a secure key (from a partially
secure key) by generating a random bit string which is completely
unknown and built using a bit string which may be partially leaked
to outside. Privacy amplification is a necessary procedure in all
QKD protocols}, where the error corrected keys are compressed by a factor determined
by CR. Now, both communicating parties would have a secure key whose
KR can be expressed as

\begin{eqnarray*}
\text{KR} & = & \text{Effective Clicks\ensuremath{\times\text{(1-DR)\ensuremath{\times\text{(1-CR)}}}}}
\end{eqnarray*}
where effective clicks $=$ Total clicks $\times$ Filtering percentage
and it differs from actual total clicks because jitter causes some
false counts to occur which need to to filter out. Filtering percentage
is a system dependent parameter that determines the amount with which
total clicks filter out. Further, even after executing LDPC protocol
some erroneous cases remain uncorrected by LDPC. Those are neglected
in obtaining the above expression. Here, it may be noted that LDPC
is not unique. In fact, there are various protocols for performing
error correction. Most widely used schemes are Cascade protocol \cite{brassard1993secret},
Low Density Parity Check (LDPC) protocol \cite{gallager1962low},
and Winnow protocol \cite{buttler2003fast}. Each of them have merits
and demerits. Specifically, Cascade protocol is most popularly used,
but it involves high communication complexity and leads to a very
low throughput. In contrast, communication complexity is reduced in
the Winnow protocol but it results into introduction of new errors.
Such errors also appear in Cascade. Interestingly, LDPC protocol can
reasonably reconcile such errors at a speed higher than that in the
Cascade or Winnow scheme. However, its computational complexity is
high \cite{johnson2015analysis}. Despite the fact that LDPC has higher
computational complexity, here we have used this protocol because
it can reconcile errors to a reasonable extent.

\section{Experimental observations\label{sec:Observation}}

The performance of any QKD protocol is quantified in terms of the
secret KR which may be computed by the process described in the previous
section. This quantity is computed with the help of relevant experimental
parameters. In the experimental setup described above, initial data
for KR calculation has been taken by varying different parameters
like DR, CR, DT and BV of the detector at various distances. Before
we proceed further, it would be apt to note DR refers to the procedure
in which Alice and Bob take a fraction of the key and compare it publically
to check errors. whereas CR refers to the amount with which error
corrected key has to be compressed in order to enhance the security
of the key and BV is the difference between SPAD bias voltage and
its breakdown voltage. Now to begin our investigation on the impact
of variation of different experimental parameters on the KR we may
first illustrate variation of KR with DR.

\subsection{Variation of key rate with disclose rate}

\begin{figure}
\begin{centering}
\begin{tabular}{ccc}
\includegraphics[scale=0.5]{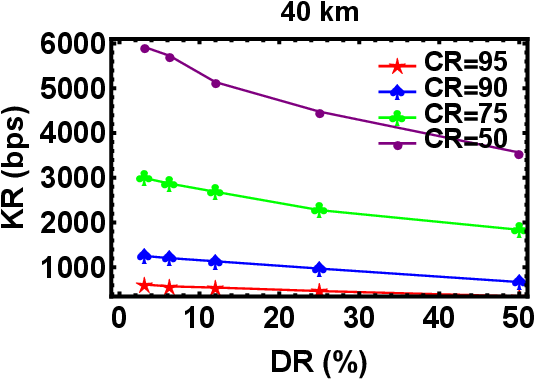} & \includegraphics[scale=0.5]{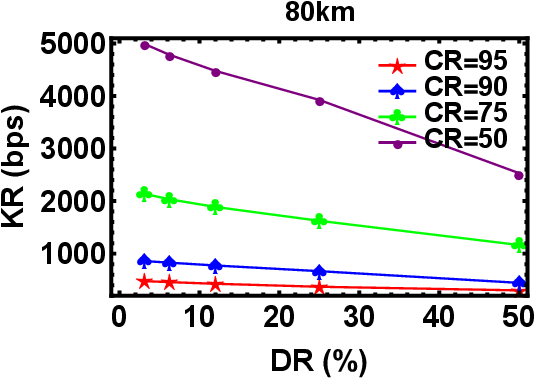} & \includegraphics[scale=0.5]{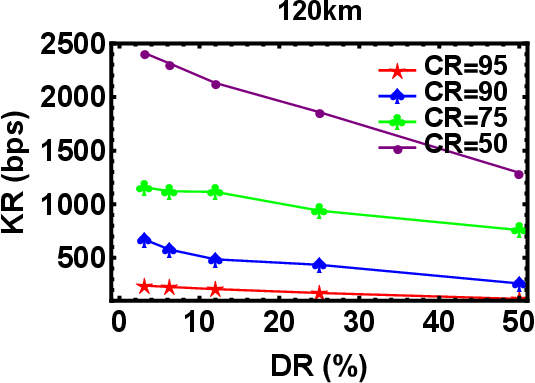}\tabularnewline
(a) & (b) & (c)\tabularnewline
\includegraphics[scale=0.5]{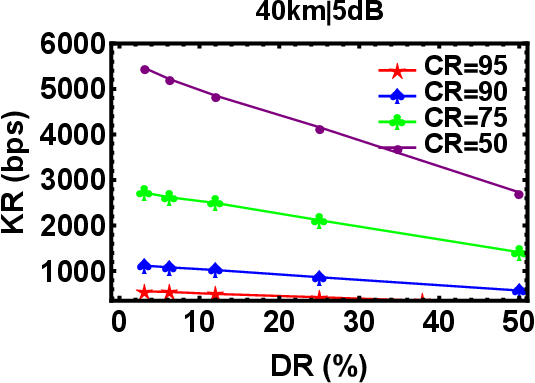} & \includegraphics[scale=0.5]{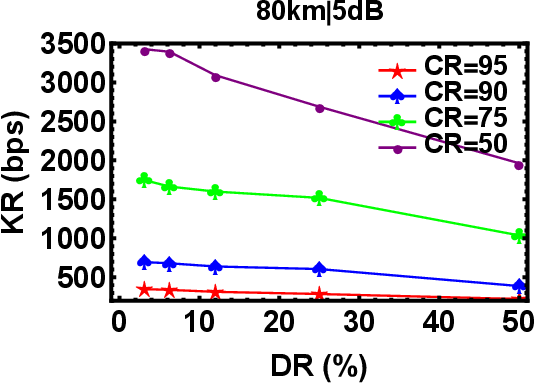} & \includegraphics[scale=0.5]{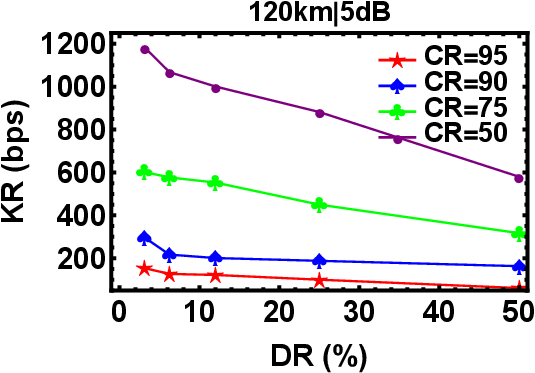}\tabularnewline
(d) & (e) & (f)\tabularnewline
\end{tabular}
\par\end{centering}
\caption{\label{fig:COW-DR}(Color Online) The KR for COW protocol with respect
to DR for different CR (a) for $40$ km (b) for $80$ km (c) for $120$
km, (d) for $40\,{\rm km}+5{\rm \,dB}\sim65\,{\rm km}$ (e) for $80\,{\rm km}+5{\rm \,dB}\sim105\,{\rm km}$
(f) for $120\,{\rm km}+5{\rm \,dB}\sim145\,{\rm km}$, keeping DT
$=50\mu$s and BV $=2$V. KR is decreasing with increase in distance
and DR. }
\end{figure}

The plots shown in Fig. \ref{fig:COW-DR} (a), (b) and (c) are for
variation KR as a function of DR at different CR for various distances
$40$, $80$ and $120$ km, respectively. The plots shown in Fig.
\ref{fig:COW-DR} (d), (e) and (f) are also for KR as a function of
DR but with introducing a loss of 5 dB using an attenuator in the
quantum channel, at different CR for various distances $40$, $80$
and $120$ Km, respectively. It can be clearly observed that there
is a linear increase in KR as we decrease the DR from $\text{50\%}$
to $\text{3.125\%}$, but a slight variation in the linear nature
of KR can be observed when we decrease DR below $\text{\ensuremath{10} \%}$
for distance $\geq120$ km (as reflected through Fig. \ref{fig:COW-DR}
(c) and (f)). Here, it may be noted that the relevance of DR is much
more in DV protocols like BB84, where DR of $50$\% is required for
achieving asymptotic security against eavesdropping \cite{nielsen2010quantum}.
In COW protocol, eavesdropping is checked through monitoring line
and DR does not play the same role as in BB84. Thus, after checking
consistency of QBER one can perform the rest of the experiment with
a lower value of DR. Our setup allows us to control DR and we have
performed experiment up to $50$\% DR and checked that QBER does not
change appreciably if we change DR in the range $3.125$\% to $50$\%.
Keeping this in mind, maximum achievable KR in COW QKD for a fixed
distance, DT and BV can be computed for DR of $3.125$\%. Though Fig.
\ref{fig:COW-DR} intrinsically illustrates the variation of KR with
CR and distance, too (along with variation with DR). We can explicitly
check the variation of KR with CR.

\subsection{Variation of key rate with compression ratio}

\begin{figure}
\begin{centering}
\begin{tabular}{ccc}
\includegraphics[scale=0.5]{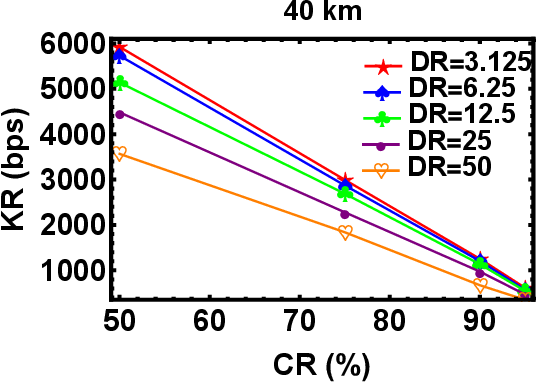} & \includegraphics[scale=0.5]{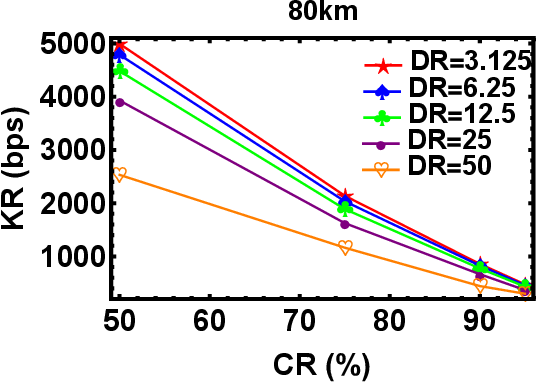} & \includegraphics[scale=0.5]{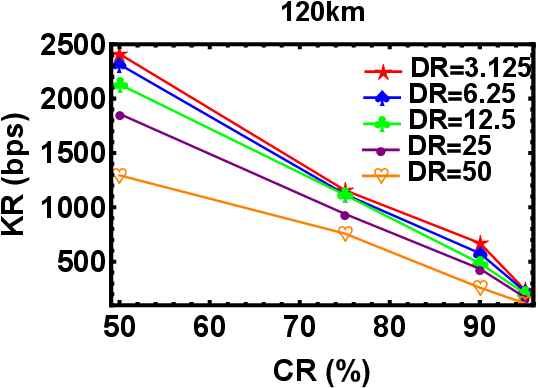}\tabularnewline
(a) & (b) & (c)\tabularnewline
\includegraphics[scale=0.5]{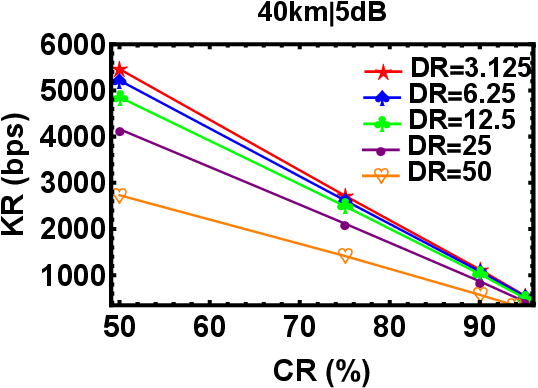} & \includegraphics[scale=0.5]{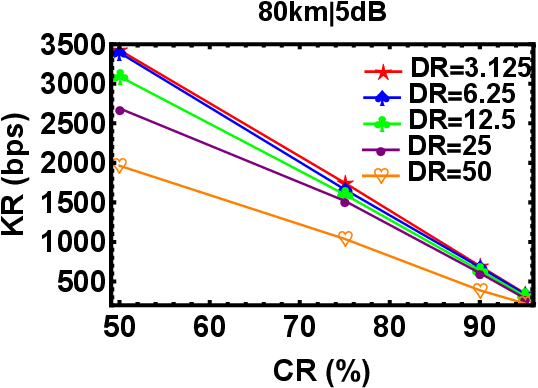} & \includegraphics[scale=0.5]{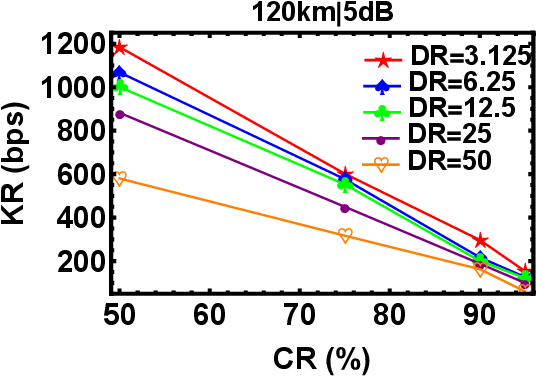}\tabularnewline
(d) & (e) & (f)\tabularnewline
\end{tabular}
\par\end{centering}
\caption{\label{fig:COW-PA}(Color Online) The KR for COW protocol with respect
to CR for different DR (a) for $40$ km (b) for $80$ km (c) for $120$
km, (d) for $40\,{\rm km}+5{\rm \,dB}\sim65\,{\rm km}$ (e) for $80\,{\rm km}+5{\rm \,dB}\sim105\,{\rm km}$
(f) for $120\,{\rm km}+5{\rm \,dB}\sim145\,{\rm km}$, keeping DT
$=50\mu$s and BV $=2$V. KR is decreasing with increase in distance
and increasing with DR. }
\end{figure}

The plots shown in Fig. \ref{fig:COW-PA} (a), (b) and (c) are for
KR as a function of CR at different DR for distances $40$, $80$
and $120$ km respectively. The plots shown in Fig. \ref{fig:COW-PA}
(d), (e) and (f) are also for KR as a function of CR with $\text{\ensuremath{5} dB}$loss
in the quantum channel at different DR for various distances $40$,
$80$ and $120$ km, respectively. It is to be noted that our system
is programmed in such a way that DR can take only certain values,
and in the previous subsection we have already mentioned that it's
sufficient to keep DR at $3.125$\%. So keeping DR at $3.125$\%,
we can compute the maximum possible KR for each case illustrated here.
Here, we investigate the variation of KR with CR. It is observed that
there is a linear increase in KR as we decrease the CR from $\text{\ensuremath{95} \%}$
to $\text{\ensuremath{50} \%}$. Also, one can easily see that KR
shows its maximum value at $\text{\ensuremath{3.125} \%}$ DR for
various distances as is expected from the discussion of the previous
subsection. In fact it's quite obvious as the disclosed bits are not
used for key generation, the lesser is the DR the more is the KR.

\subsection{Variation of key rate with detector dead time}

\begin{figure}
\begin{centering}
\begin{tabular}{cc}
\includegraphics[scale=0.6]{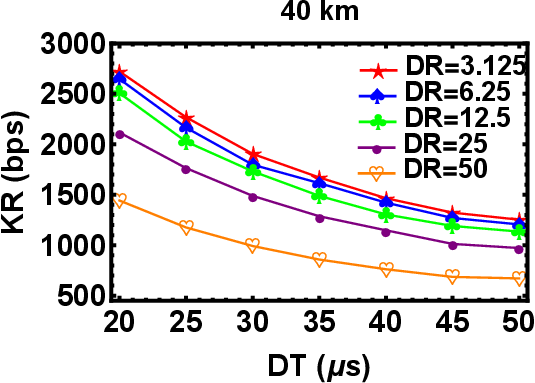} & \includegraphics[scale=0.6]{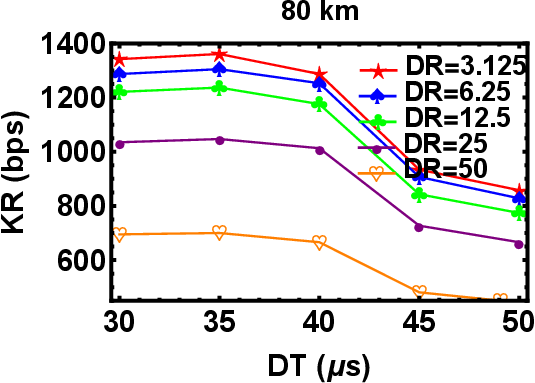}\tabularnewline
(a) & (b)\tabularnewline
\end{tabular}\caption{\label{fig:KR vs.DT_@vDR}(Color Online) The secret KR for the COW
protocol with respect to DT of the detector at different DR for (a)
$40$ km and (b) $80$ km distance keeping CR =$\,90\%$. Figures
illustrate that KR increases with the decrease in DT. In case of $80$
km distance, key rate below DT $30$ $\mu$s was not significant.
Data for $120$ km distance is not plotted for reduced DT as QBER
was observed to increase considerably for lower values of DT in that
case.}
\par\end{centering}
\end{figure}

It's expected that KR will be a function of DT and KR will decrease
with increase in DT as for an increased DT, the detector will remain
blind for a longer period and will not be able detect pulses arriving
at the detector within that period. However, this logical expectation
does not tell us the actual nature of variation of KR with DT. To
provide the same, in Fig. \ref{fig:KR vs.DT_@vDR} (a) and (b) we
illustrate the variation of KR as a function of DT of the detector
used at Bob side, at different DR and fixed CR to $90$ \% for distances
$40$ and $80$ km respectively. It is to be noted that one cannot
decrease DT of the detector below the time gap between two consecutive
pulses coming from Alice's side. Also, with a decrease in DT, the
dark count rate (DCR) of the detector increases. The detector used
here is set at a DT of 50 $\mu s$ by default. Here, in the Fig. \ref{fig:KR vs.DT_@vDR},
the secure KR is continuously increasing with decrease in the DT for
$40$ km distance while one can see an abrupt change when the DT goes
down from $45$ ${\rm \mu s}$ to $40$ ${\rm \mu s}$ for $80$ km
distance. This abrupt change may be attributed to the sudden increase
in DCR of the detector. So, it is clear that one cannot extract secure
KR by going below 45 ${\rm \mu s}$ DT for $80$ km of secure quantum
communication implemented using the present approach and the detectors
used here.

\begin{figure}
\begin{centering}
\begin{tabular}{cc}
\includegraphics[scale=0.6]{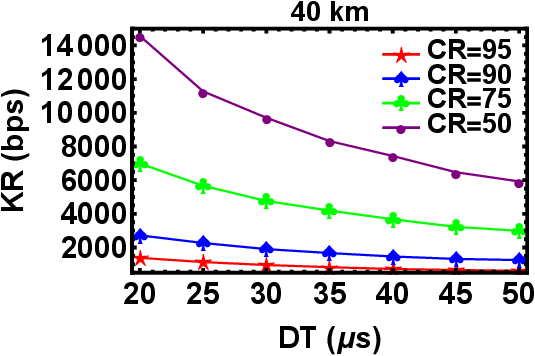} & \includegraphics[scale=0.6]{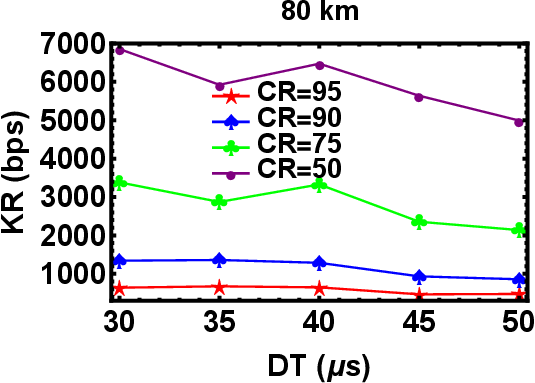}\tabularnewline
(a) & (b)\tabularnewline
\end{tabular}\caption{\label{fig:KR vs.DT_@vCR}(Color Online) The KR for the COW protocol
with respect to DT of the detector at different CR for (a) $40$ km
and (b) $80$ km distance keeping DR =$\,3.125\,\%$. The KR increases
with decrease in DT. In case of $80$ km distance KR below DT $30$
$\mu$s was not significant. }
\par\end{centering}
\end{figure}

Above was an illustration of variation of KR with DT for a very high
CR. However, such a high CR may not be required\footnote{For example in Ref. \cite{walenta2014COW} CR of $6.5$\% is used.
As secret KR is computed simply multiplying the sifted KR by CR \cite{tisi2021multiplexing},
reduction of CR to such a value will considerably enhance the KR reported
here. } Keeping that in mind, and in an effort to provide completeness to
this study, in Fig. \ref{fig:KR vs.DT_@vCR} (a) and (b) we plot KR
as a function of DT of the detector for different values of CR and
fixed DR to $3.125
$ for distances $40$ and $80$ km, respectively. Here, the secure
KR is found to slightly increase with the decrease in DT at CR $90
$ and $95
$ for $40$ km distance while the secure KR is observed to exponentially
increase at CR $75
$ and $50
$ but the less we compress the key, the less secure will be the key.
Hence, to increase the security of the key, it is better to have a
reasonably high CR which may computed using the method described in
\cite{tisi2021multiplexing}. Further, it may be noted that in Ref.
\cite{walenta2014COW}, secure KR for COW QKD protocol is computed
for a CR of $6.5\%$ for $25$ km distance. We may further note that
an abrupt change in KR can also be seen when DT goes from $45$ ${\rm \mu s}$
to $40$ ${\rm \mu s}$ for $80$ km of length of the quantum channel.
So, once again it is clear that one cannot extract secure KR by going
below $45$ ${\rm \mu s}$ DT for $80$ km quantum communication distance
with the hardware used here.

Finally, we must note that we have already seen that secure key generation
in a dependable and consistent manner is not possible at DT below
$45$ ${\rm \mu s}$ for $80$ km long quantum channel for the COW
protocol for QKD implemented here. For 120 km of fiber optic quantum
channel, it's observed that secure KR not obtained for DT $<50$ ${\rm \mu s}$.
This is why we have not attempted to show variation of KR with DT
for $120$ km or $120$ km+$5$ dB loss. Of course, for DT of $50$
${\rm \mu s}$, we can also study the variation of KR with CR and
DR for $120$ km and $120$ km+$5$ dB loss. The same is already done
and shown in Fig. \ref{fig:COW-DR} (c) and (f) and \ref{fig:COW-PA}
(c) and (f), respectively. 

\begin{figure}
\begin{centering}
\begin{tabular}{cc}
\includegraphics[scale=0.6]{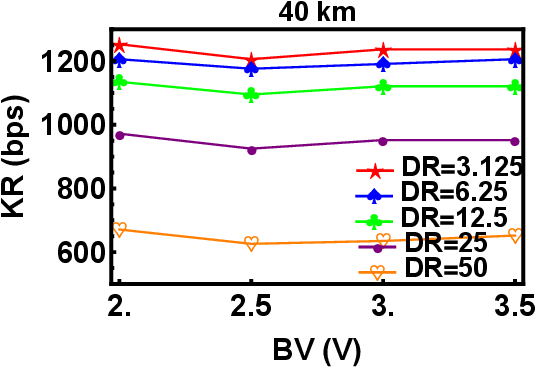} & \includegraphics[scale=0.6]{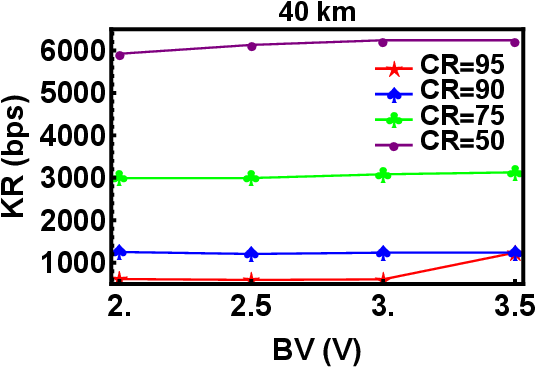}\tabularnewline
(a) & (b)\tabularnewline
\end{tabular}
\par\end{centering}
\caption{\label{fig:KR vs.BV_40km}(Color Online) The KR for the COW protocol
with respect to BV of the detector at different (a) DR keeping CR
$90\%$and (b) CR keeping DR $3.125\%$ for $40$ km (c) KR with respect
to time.}

\end{figure}

\subsection{Variation of key rate with excess bias voltage }

The plots shown in Fig. \ref{fig:KR vs.BV_40km} (a) and (b) are for
KR as a function of BV of the detector at different DR and different
CR with fixed CR $90$\% and fixed DR $3.125\ensuremath{\%}$, respectively
for quantum communication distances of $40$ km. Here, the secure
KR is almost constant with change in BV which that indicates there
is no significant impact of BV on KR. This is because the change in
order of the DCR is almost negligible with change in BV up to DT of
the detector become $20$ ${\rm \mu s}$. However, this conclusion
is strictly valid for $40$ km only. For larger distance, DCR is not
negligible in the similar situation.

\subsection{Variation of key rate with time}

\begin{figure}
\begin{centering}
\includegraphics[scale=0.5]{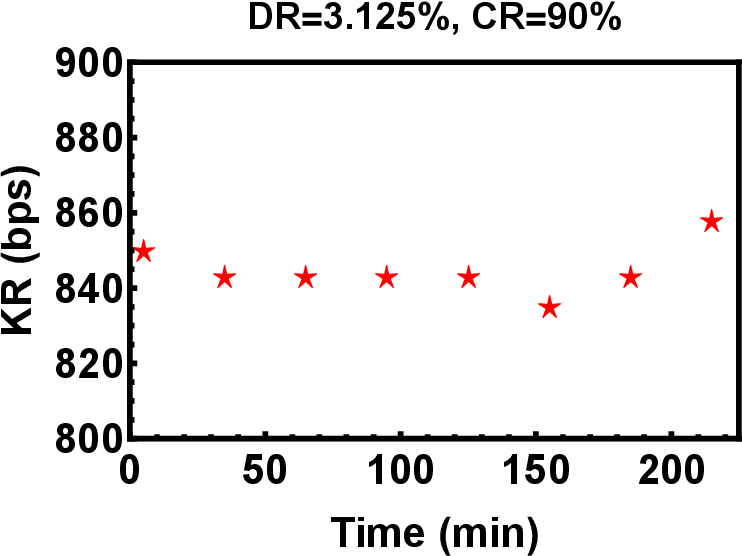}
\par\end{centering}
\caption{\label{fig:key-rate-with-time}(Color Online) The KR is found to remain
stable with time. This plot is for 80 km of optical fiber.}
\end{figure}

Stability of KR is very important. It's required for finite key analysis
\cite{walenta2014COW} and also for checking the stability of the
system. Here in analogy with several existing works \cite{Gisin2009COW}
we have checked the stability of KR for two hours as shown in Fig.
\ref{fig:key-rate-with-time}. It's found that the KR remains stable
in our system. Here, we have shown the time variation of KR for 80
km of quantum channel, but similar characteristics are observed for
40 km and 120 km, too when DR=$3.125
$, CR=$90\%$, DT=$50\mu$s, and BV=$2$V are selected as experimental
parameters. 

\section{Comparison of the experimental implementations of the COW protocol\label{sec:Comparison}}

While describing our experimental observations, we have briefly mentioned
the differences of the present implementation with the existing implementations
with specific focus on the limitations and advantages of the present
work. Here, we aim to perform the comparison in a more systematic
manner. Aiming that, comparison of our work with the existing implementations
of COW protocol is summarized in Table \ref{tab:Literature survey}.
In fact, Table \ref{tab:Literature survey} summarizes the experimental
progress that has happened so far in the implementation of the COW
protocol for QKD. As all the implementations are fiber based, that
aspect is not specifically mentioned in the table. 

\begin{table}
\begin{centering}
\begin{tabular}{|>{\centering}p{1.8cm}|>{\centering}p{1.5cm}|>{\centering}p{1.8cm}|>{\centering}p{2cm}|>{\centering}p{2cm}|>{\centering}p{1.5cm}|>{\centering}p{3cm}|>{\centering}p{1.5cm}|}
\hline 
Wavelength

(nm) & Nature of RNG used & Error correction & PA scheme & Detector Type & Distance (km) & KR (bps) & References\tabularnewline
\hline 
$1550$ & TRNG & LDPC & FFT based Toeplitz hashing scheme & InGaAs SPD (free running mode) & (a)120

(b) 145 & (a) $241-2410$

(b)$154-1184$ & Our work\tabularnewline
\hline 
$1550$  & QRNG to seed PRNG & Cascade & Hashing functions based on Toeplitz matrices & InGaAs SPD (Lab) & $150$ & $>50$ & \cite{stucki2009continuous}\tabularnewline
\hline 
$1550$  & QRNG to seed PRNG & Cascade & Hashing functions based on Toeplitz matrices & SNSPD (Field) & $150$ & $2.5$ & \cite{stucki2009continuous}\tabularnewline
\hline 
$1550$  & QRNG & Cascade & Hashing functions based on Toeplitz matrices & SNSPD & $100-250$ & $6000-15$ & \cite{Gisin2009COW}\tabularnewline
\hline 
$1310$  & QRNG & \# & \# & InGaAs SPD & (a) $14.2$

(b) $45.6$

(c) $24.2$

(d) $36.6$

(total 121 km with three trusted nodes) & (a) $1956$

(b) $1314$

(c) $763$

(d) $1906$ & \cite{121kmCOW}\tabularnewline
\hline 
$1550$  & \textcolor{black}{QRNG} & Cascade & Universal hash functions & InGaAs/InP negative feedback avalanche diodes (NFADs) & $307$ & $3.18$ & \cite{korzh2015COW}\tabularnewline
\hline 
$1551.72$  & QRNG to seed PRNG & LDPC & Hashing functions based on Toeplitz matrices & InGaAs SPD & $25$ & $22500$ & \cite{walenta2014COW}\tabularnewline
\hline 
$1550.12$ & \# & LDPC & Hashing functions based on Toeplitz matrices & InGaAs SPD (gated mode) & $150$ & $<1000$ & \cite{shaw2022optimal}\tabularnewline
\hline 
\end{tabular}
\par\end{centering}
\caption{\label{tab:Literature survey} Experimental developments for the fiber
based COW-QKD protocol. \# means that the information is not provided
clearly in the mentioned paper.}
\end{table}

In Table \ref{tab:Literature survey}, we can see that both Cascade
and LDPC protocols for error correction are used. Different protocol
for error correction is used in different experiments. Here, we have
used the LDPC error correction scheme, which is also used in \cite{walenta2014COW}
and in \cite{shaw2022optimal}. We have changed the LDPC code rate
and performed the experiment. However, in contrast to Ref. \cite{walenta2014COW}
where it was found that the best KR is obtained for LDPC code rate
of $3/4$ we have not found any appreciable change in secure KR for
the experimental conditions used in this work. Interestingly, quantum
LDPC \cite{gottesman2014fault,breuckmann2021quantum} has drawn considerable
attention of the community in the recent past, in near future quantum
LDPC may be used for error correction in the experimental realizations
of COW protocol for QKD.

In most of the published works, post processing parameters are not
mentioned, and consequently, it's not possible to directly compare
the KRobtained in their implementations and that obtained in our case.
So we have decided to compare the worst result of our implementation
with the existing implementations. The worst result is obtained for
a very high CR (CR of $90$\%). Definitely, this ensures security,
but such a high CR is not always needed as we have discussed above.
Further, in \cite{walenta2014COW} for a distance of $25$ km a KR
of $22500$ bps was reported for CR of $6.5$\%. If one desires to
perform the same experiment with CR=$90$\%, then the KR would become
$\sim1562$ bps only, which is not better than ours realization as
for 40 km distance (which is higher than $25$ km) for CR=$50$\%
we have obtained KR=$5923$ bps for DR=$3.125$\% and DT=$50$ $\mu s$
and the same was as much as $14562$ bps for DT=20 $\mu s$ while
other parameters are kept the same. Thus, comparison among other works
are difficult in absence of post possessing parameters. Now, for such
a high value of CR (CR=$90$\%) we obtain maximum secure KR of $1466$
bps, $934$ bps and $671$ bps for $40$ km, $80$ km and $120$ km
respectively at DT 40 $\mu s$, 45 $\mu s$ and 50 $\mu s,$ respectively
by keeping DR value fixed to $3.12$5 \%. Of course, for lower values
of CR we get much better secure KR. For example, if we consider CR=$50$\%,
DR=$3.125$\% then the obtained secure KR would become $14562$ bps
(for $40$ km and $20$~$\mu$s DT), $6853$ bps \footnote{An abrupt change is observed when DT goes down from $45\text{\,\ensuremath{\mu s}}$
to $40\,\mu s$ due to increase in the DCR of the detector. So, the
obtained KR is $5637$ bps for DT $45\,\mu s$} (for $80$ km and $30$$\,\mu$s DT), $2410$ bps (for $120$ km
and $50\,$$\mu$s DT) for the lowest value of DT of detector allowed
in our hardware for the specific cases. We have not tried CR <$50$\%,
so our KR varies in the range $616-14562$ bps for $40$ km, in the
range $476-6853$ bps for 80 km and in the range $241-2410$ bps for
$120$ km and in the range $154-1184$ bps for $145$ km (where in
the last case $25$ km of fiber is emulated by using an attenuator
of $5$ dB loss) for different choices of experimental parameters.
However, even better secure KR can be obtained by choosing lower values
of KR as was done in \cite{walenta2014COW}.

Now, we would like to note that even in the worst case scenario, we
are obtaining secure KR comparable to other recent implementations.
For example, in 2023, in Ref. \cite{121kmCOW}, COW QKD was implemented
for a distance of 121 km using three trusted nodes. We have implemented
COW QKD for almost similar distance (exactly 120 km) without any trusted
node. They obtained secure KR in the range, 763 bps-1956 bps as the
distance between different nodes were different and obtained KR was
also different. However, a more involved theoretical analysis\textcolor{red}{{}
\cite{duan2023COW} }using the experimental data of \cite{121kmCOW}
has yielded a secure KR of $790$ bps for a distance of $121$ km.
Now, even in the worst case scenario, we have obtained $671$ bps
for almost similar distance without using any trusted node. 

Finally, the distance over which COW protocol for QKD is realized
here is higher than that in \cite{walenta2014COW} and of the order
of the implementations reported in Refs. \cite{stucki2009continuous,121kmCOW,duan2023COW,shaw2022optimal}
obtained secure KR are of the similar order. Further, it's important
to note that most of these experiments reporting the implementation
of COW QKD for distances in the range $121-150$ km happened in the
last one year (see \cite{121kmCOW,duan2023COW,shaw2022optimal}).
Thus, the present experimental realization is apparently consistent
with the state-of-the-art experiments. However, there are reports
where COW QKD is implemented over larger distances. For example, see
Refs. \cite{korzh2015COW,Gisin2009COW}. In these works COW QKD protocol
is implemented for very large distances. For example, in \cite{Gisin2009COW}
it's implemented for distances up to $250$ km, but to do so SNSPD
and ultra low loss fibers were used and a very small KR ($15$ bps)
was observed for $250$ km distance. Similarly, in \cite{korzh2015COW},
COW protocol for QKD is realized for $307$ km using ultra low loss
optical fiber and a KR of $3.18$ bps is obtained. From this comparison,
we may conclude that the KR reported here is reasonable compared to
earlier results, but there are several known avenues for enhancing
the same. Specifically, standard telecom grade optical fiber is used
here which has $0.2$ dB/km loss, if we replace that with ultra low
loss optical fiber (Corning$^{\circledR}$ SMF-28$^{\circledR}$ULL
fiber) having $0.164$ dB/km loss (as was used in \cite{Gisin2009COW})
and the other fiber (ultra low loss silica based fiber) of loss $0.160$
dB/km (used in \cite{korzh2015COW}), KR and/or distance can be increased
substantially. Further, the use of better detectors like SNPSDs or
InGaAs SPDs with relatively low background noise can improve the KR.
Definitely the impact of using SNSPDs on KR will be higher than that
of using better quality InGaAs SPDs.

\section{Conclusions\label{sec:Conclusions}}

The COW QKD protocol is one of the most practical protocols in the
domain of quantum cryptography. As described above, in this protocol,
there are two participants, Alice and Bob. Alice, the sender, sends
pulses and Bob detects those pulses. In her lab, Alice has a laser
source and attenuator followed by an intensity modulator and Bob has
optical coupler and detectors in his lab. Alice prepares a random
sequence of either a pulse of mean photon number $\mu\,(\mu=\left|\alpha\right|^{2})$
or a vacuum pulse and sends to Bob. While receiving the pulse sequence,
Bob measures the time of arrival of $90\%$ photons at his detector
for the generation of raw key and rest $10\%$ of photons on monitoring
line for security purpose. In this article, we report a lab level
experimental realization of COW QKD protocol for various effective
distances in the range $40$ km to $145$ km. We also report variation
of secure KR with various experimental parameters which are not usually
discussed in detail. However, a good understanding of these variations
and hardware limitations (specially the limitations of the detector)
can be of much use for the future experiments and commercial deployment
of COW QKD schemes. Specifically, variation of KR is observed with
DT, DR, CR, BV, etc., and efforts have been made to understand the
physical origin of fluctuations wherever observed. In the entire experiment,
QBER is always kept below $6$\%. 

This experiment shows how to implement COW QKD in the lab without
a monitoring line and analyze the system for various experimental
parameters. In near future several generalizations of this experiment
are planned. Specifically, a realization with monitoring line will
be reported soon. Subsequently, the experiment will be performed with
SNSPD, and efforts will be made to enhance the distance by using ultra
low loss fibers as was done in \cite{Gisin2009COW,korzh2015COW}.
Further, zero error attack on COW protocol may be attempted. Zero
error attacks are those attacks in which Eve does not incorporate
any error still the authorized parties i.e., Alice and Bob are unable
to form a secret key. It is a kind of sequential attack in which Eve
first of all measures all the bits and whenever she obtains the conclusive
results she sends corresponding bits to Bob otherwise she sends vacuum
to Bob. In case of COW protocol, this attack would utilize two facts:
(i) in the COW protocol vacuum state is used, (ii) Alice's states
are linearly independent. 

Keeping all the above possibilities of further generalization of the
present experiment in mind, we conclude this article in an optimistic
mode that the physical insights obtained here and the simple approach
followed here will lead to a set of new and exciting experimental
studies on DPR class of QKD protocols with specific focus on COW protocol.

\section*{Acknowledgments}

Authors acknowledge the support from the QUEST scheme of Interdisciplinary
Cyber Physical Systems (ICPS) program of the Department of Science
and Technology (DST), India (Grant No.: DST/ICPS/QuST/Theme-1/2019/14
(Q80)) and the entire CDOT team working on quantum technologies. They
also thank Sandeep Mishra, Kishore Thapliyal and Abhishek Shukla for
their interest in the work.

\section*{Availability of data and materials}

Experimental data will be made available on request.

\section*{Competing interests}

The authors declare that they have no competing interests.

\bibliographystyle{unsrt}
\bibliography{COW-paper-final-draft1.bib}

\end{document}